\documentclass[
superscriptaddress,
reprint,
floatfix,
amsmath,amssymb,
aip, 
]{revtex4-2}

\usepackage{graphicx}
\usepackage{dcolumn}
\usepackage{bm}
\usepackage[utf8]{inputenc}
\usepackage{multirow}
\usepackage{makecell}
\usepackage{tabularx}
\usepackage{amsmath}
\usepackage{hyperref}
\usepackage{xcolor}
\begin{document}

\title{Integrated photon-pair sources on periodically poled thin-film lithium tantalate}

\author{Sakthi Sanjeev Mohanraj} \thanks{These authors contributed equally.}
\affiliation{Department of Materials Science and Engineering, National University of Singapore, Singapore 117575, Singapore}
\affiliation{Centre for Quantum Technologies, National University of Singapore, Singapore 117543, Singapore}

\author{Xiaodong Shi} \thanks{These authors contributed equally.}
\affiliation{A$^\ast$STAR Quantum Innovation Centre (Q.InC), Agency for Science, Technology and Research (A$^\ast$STAR), Singapore 138634, Singapore}

\author{Ran Yang}
\affiliation{Centre for Quantum Technologies, National University of Singapore, Singapore 117543, Singapore}

\author{Lin Zhou}
\affiliation{Department of Materials Science and Engineering, National University of Singapore, Singapore 117575, Singapore}
\affiliation{Centre for Quantum Technologies, National University of Singapore, Singapore 117543, Singapore}

\author{Di Zhu}
\email{dizhu@nus.edu.sg}
\affiliation{Department of Materials Science and Engineering, National University of Singapore, Singapore 117575, Singapore}
\affiliation{Centre for Quantum Technologies, National University of Singapore, Singapore 117543, Singapore}
\affiliation{A$^\ast$STAR Quantum Innovation Centre (Q.InC), Agency for Science, Technology and Research (A$^\ast$STAR), Singapore 138634, Singapore}

\begin{abstract}

Chip-integrated photon-pair sources based on spontaneous parametric down-conversion (SPDC) have emerged as a promising solution for scalable quantum light generation. Thin-film lithium tantalate (TFLT) is a compelling $\chi^{(2)}$ platform, combining strong nonlinearity with a high optical-damage threshold, weak photorefractive response, and ferroelectricity that enables quasi-phase matching. However, SPDC-based photon-pair generation on TFLT has not yet been demonstrated. Here, we combine high-quality periodic poling with low-loss nanophotonic waveguides to realize photon-pair sources on TFLT in both traveling-wave and resonant configurations. In periodically poled straight waveguides, we achieve broadband photon-pair generation with high efficiency (2.1~GHz~mW$^{-1}$) and coincidence-to-accidental ratio (up to $3.8\times10^{5}$). We further confirm high-purity single-photon operation via heralded second-order correlation ($g^{(2)}_\mathrm{H}(0) = 0.0018 \pm 0.0002$) and high-fidelity time-energy entanglement through Franson interference (visibility of $98.9 \pm 0.5\%$). In periodically poled racetrack resonators, we map out a broad quantum frequency comb spanning the telecom C- and L-bands. By isolating individual frequency-correlated pairs, we measure a high spectral brightness of 11~GHz~mW$^{-1}$~GHz$^{-1}$. These results are competitive with the state of the art across $\chi^{(2)}$ integrated platforms, positioning TFLT as a strong contender for integrated quantum light sources, with applications in wavelength-multiplexed quantum communications and photonic quantum information processing.

\end{abstract}

\maketitle

\section{Introduction}
Entangled photon pairs are key resources for photonic quantum technologies, underpinning applications ranging from secure communication \cite{gisin2007quantum,scarani2009security} and quantum networks \cite{kimble2008quantum, zheng2023multichip} to quantum computing \cite{walther2005experimental, o2007optical} and quantum-enhanced sensing \cite{pan2012multiphoton,giovannetti2004quantum,taylor2013biological}. As these systems move toward large-scale, real-world deployment, the demand for chip-integrated photon-pair sources that deliver high brightness, low noise, and broad spectral coverage becomes increasingly important---particularly for wavelength-multiplexed architectures \cite{wengerowsky2018entanglement,valencia2026large}, where spectral brightness and spectral coverage directly determine achievable event rates in each channel and number of supported channels.

Photon-pair generation based on spontaneous parametric down-conversion (SPDC) and spontaneous four-wave mixing (SFWM) provides a versatile approach to integrated quantum light sources, offering broadband operation and compatibility with photonic integrated circuits. Among these, on-chip SPDC in second-order nonlinear ($\chi^{(2)}$) platforms, such as thin-film lithium niobate (TFLN), AlN, InGaP, AlGaAs, and SiC  \cite{javid2021ultrabroadband,ma2020ultrabright,shi2024efficient,shi2025integrated,guo2017parametric,akin2024ingap,zhao2022ingap,rahmouni2024entangled,shi2025spontaneous}, is particularly attractive, owing to its high nonlinear efficiency, easy pump filtering, and low noise compared with third-order nonlinear ($\chi^{(3)}$) approaches \cite{ma2017silicon,yasui2025narrowband,guo2018generation,chen2024ultralow,li2025down}. These advantages are especially important for high-brightness operation, where noise from pump residue, Raman scattering, and multi-pair emission can significantly degrade quantum-state quality.

\begin{figure*}[t]
    \centering
    \includegraphics[width=\textwidth]{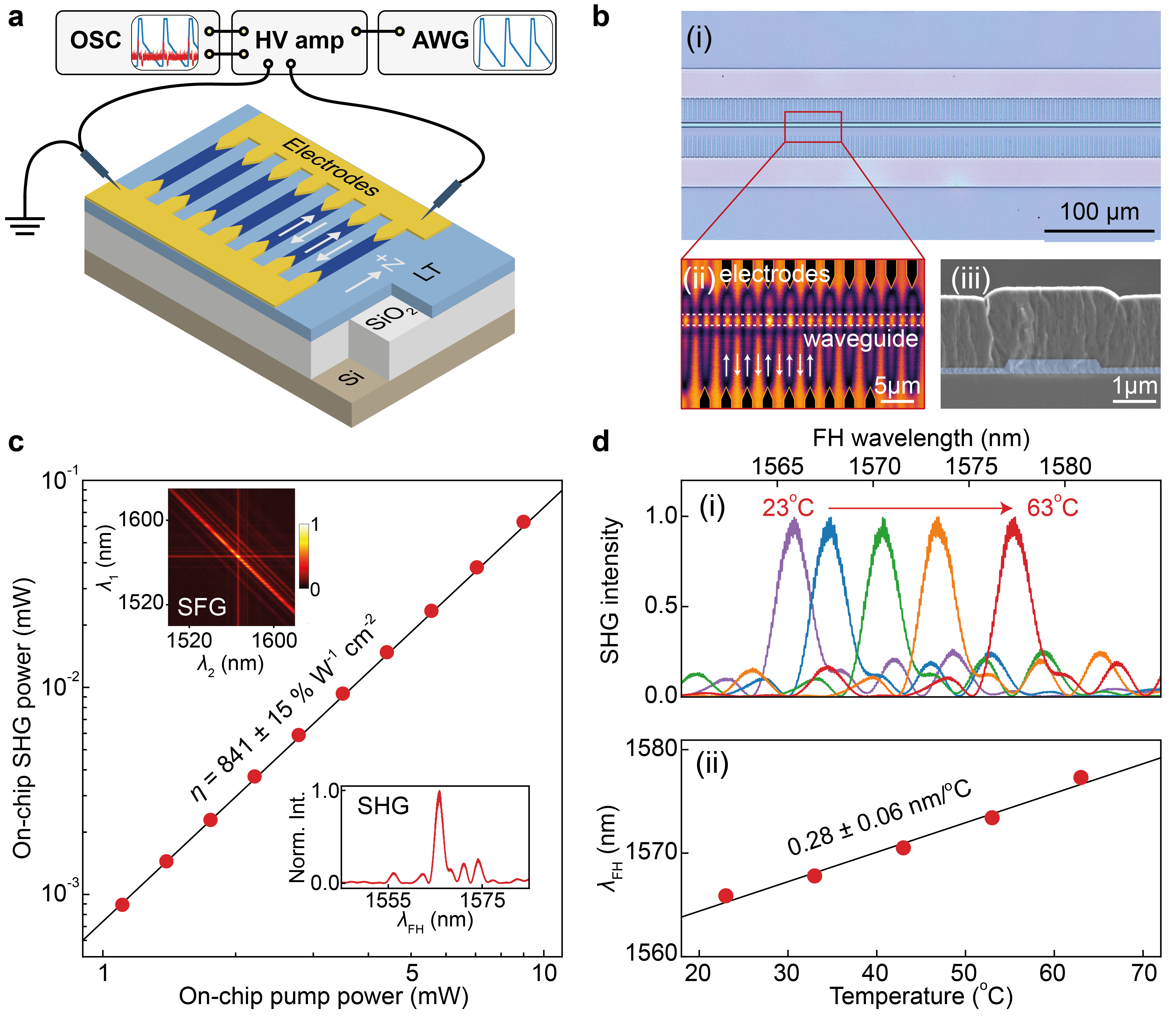}
    \caption{\textbf{Classical characterization of periodically poled lithium tantalate (PPLT) nanophotonic waveguide.} \textbf{a,} (i) Schematic illustration of the electrical poling process on thin-film lithium tantalate (TFLT). AWG: arbitrary waveform generator, HV amp: high-voltage amplifier, OSC: oscilloscope. \textbf{b,} (i) Optical micrograph of the PPLT nanophotonic waveguide, (ii) second-harmonic generation (SHG) confocal micrograph of the inverted domains after waveguide fabrication, and (iii) scanning electron micrograph (SEM) of the waveguide cross-section. \textbf{c,} On-chip SHG power as a function of pump power with normalized conversion efficiency of 841 $\pm$ 15 $\%$W$^{-1}$cm$^{-2}$ in a 3-mm-long PPLT waveguide. Insets: measured sum-frequency generation (SFG) mapping out the phase-matching function and normalized SHG spectrum. \textbf{d,} (i) Normalized SHG spectrum measured at different temperatures from 23$^{\circ}$C to 63$^{\circ}$C. (ii) Phase-matching wavelength as a function of temperature showing a red shift of 0.28 $\pm$ 0.06 nm/$^{\circ}$C with increasing temperature.}
    \label{Fig1}
\end{figure*}

Recently, lithium tantalate (LiTaO$_3$, LT) has emerged as a compelling $\chi^{(2)}$ integrated material platform, combining a broad transparency window, strong $\chi^{(2)}$ nonlinearity, and ferroelectricity that enables electric poling for quasi-phase matching (QPM). Compared with its more widely studied counterpart lithium niobate (LN), LT offers weaker photorefractive response \cite{sayem2026high}, higher optical-damage threshold \cite{wang2025thin,kuznetsov2025watt}, and lower birefringence \cite{wang2024lithium}, properties that collectively support stable operation at elevated optical powers \cite{wang2025thin} and facilitate broadband nonlinear interactions \cite{zhang2025ultrabroadband}. These advantages have driven rapid advancement in the thin-film lithium tantalate (TFLT) platform, including demonstrations of low-loss waveguides \cite{he2025lithium}, high-speed electro-optic modulators \cite{wang2024ultrabroadband,wang2025thin}, and efficient $\chi^{(2)}$ nonlinear frequency conversion \cite{chen2025continuous,yu2025efficient,shelton2025robust,kuznetsov2025watt}. Despite impressive progress in classical photonics, the platform's potential for quantum light generation remains largely unexplored. Photon pairs have only been observed via SFWM \cite{Xu2025QuantumCO}, leaving the platform's native $\chi^{(2)}$ nonlinearity and the high efficiency and low noise it affords through SPDC, unexploited. To date, photon-pair generation based on SPDC has not been demonstrated in TFLT.

Here, we demonstrate SPDC photon-pair generation in TFLT by combining high-quality periodic poling with low-loss waveguides and racetrack resonators.
In straight periodically poled lithium tantalate (PPLT) waveguides, we realize efficient SPDC with pair generation efficiency exceeding 2.1 GHz mW$^{-1}$ and coincidence-to-accidental ratio (CAR) of 3.8 $\times$ 10$^5$. 
We verify high-fidelity time-energy entanglement using Franson-type two-photon interferometry with interference visibility of 98.9$\pm0.5\%$. 
In PPLT racetrack resonators, we generate bi-photon quantum frequency comb (QFC) covering the telecom C- and L-bands, achieving spectral brightness as high as 11 GHz mW$^{-1}$ GHz$^{-1}$ and a CAR of up to 2.8 $\times$ 10$^3$. This is among the best-performing integrated photon-pair sources in terms of spectral brightness and is naturally suited for dense wavelength division multiplexing (DWDM).
Our results establish TFLT as a promising platform for integrated quantum photonics.

\begin{figure*}[t]
    \centering
    \includegraphics[width=6.3in]{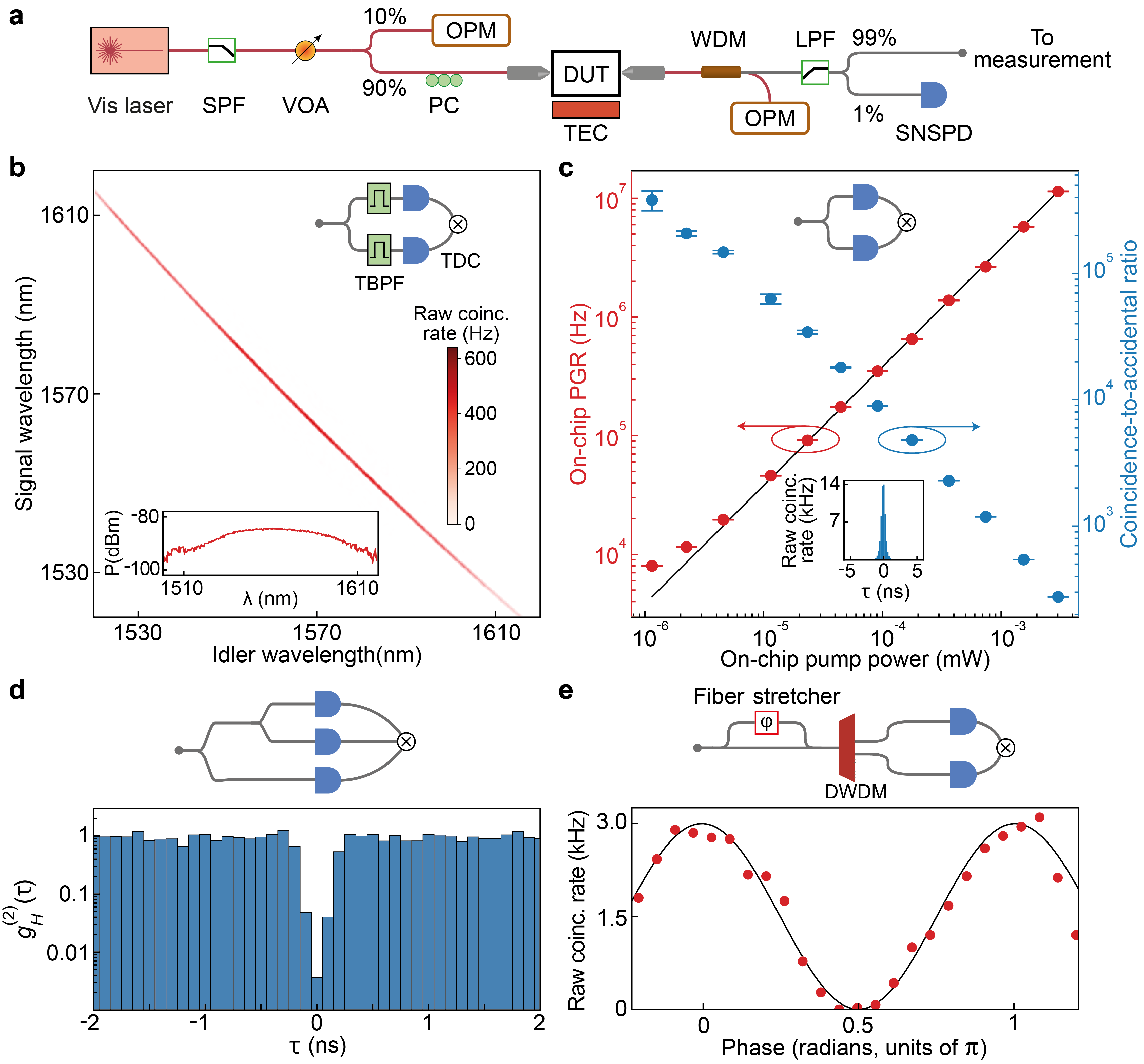}
    \caption{\textbf{Nonclassical characterization of PPLT nanophotonic waveguide.} \textbf{a,} Measurement setup for photon-pair generation. Vis laser: visible laser, SPF: short-pass filter, VOA: variable optical attenuator, OPM: optical power meter, PC: polarization controller, DUT: device under test, TEC: thermoelectric cooler, WDM: wavelength division multiplexer, LPF: long-pass filter, SNSPD: superconducting nanowire single-photon detector. \textbf{b,} Joint spectral intensity of the photon pairs with insets of the detection scheme (top) and SPDC spectrum obtained by optical spectrum analyzer (bottom). \textbf{c,} On-chip photon-pair generation rate (PGR) and coincidence-to-accidental ratio (CAR) as a function of pump power, with insets of the detection scheme (top) and a recorded coincidence histogram at the highest PGR (bottom). \textbf{d,} Heralded second-order correlation detection scheme along with the self correlation as a function of time delay, resulting in an anti-bunching dip $g_\mathrm{H}^{(2)}(\tau)$ of 0.0018 $\pm$ 0.0002. \textbf{e,} Folded Franson-type two-photon interference detection scheme and the interference observed as a function of phase shift, showing a raw visibility of 98.9$\pm0.5\%$ without accidental subtraction.}
    \label{Fig2}
\end{figure*}

\begin{figure*}[t]
    \centering
    \includegraphics[width=\textwidth]{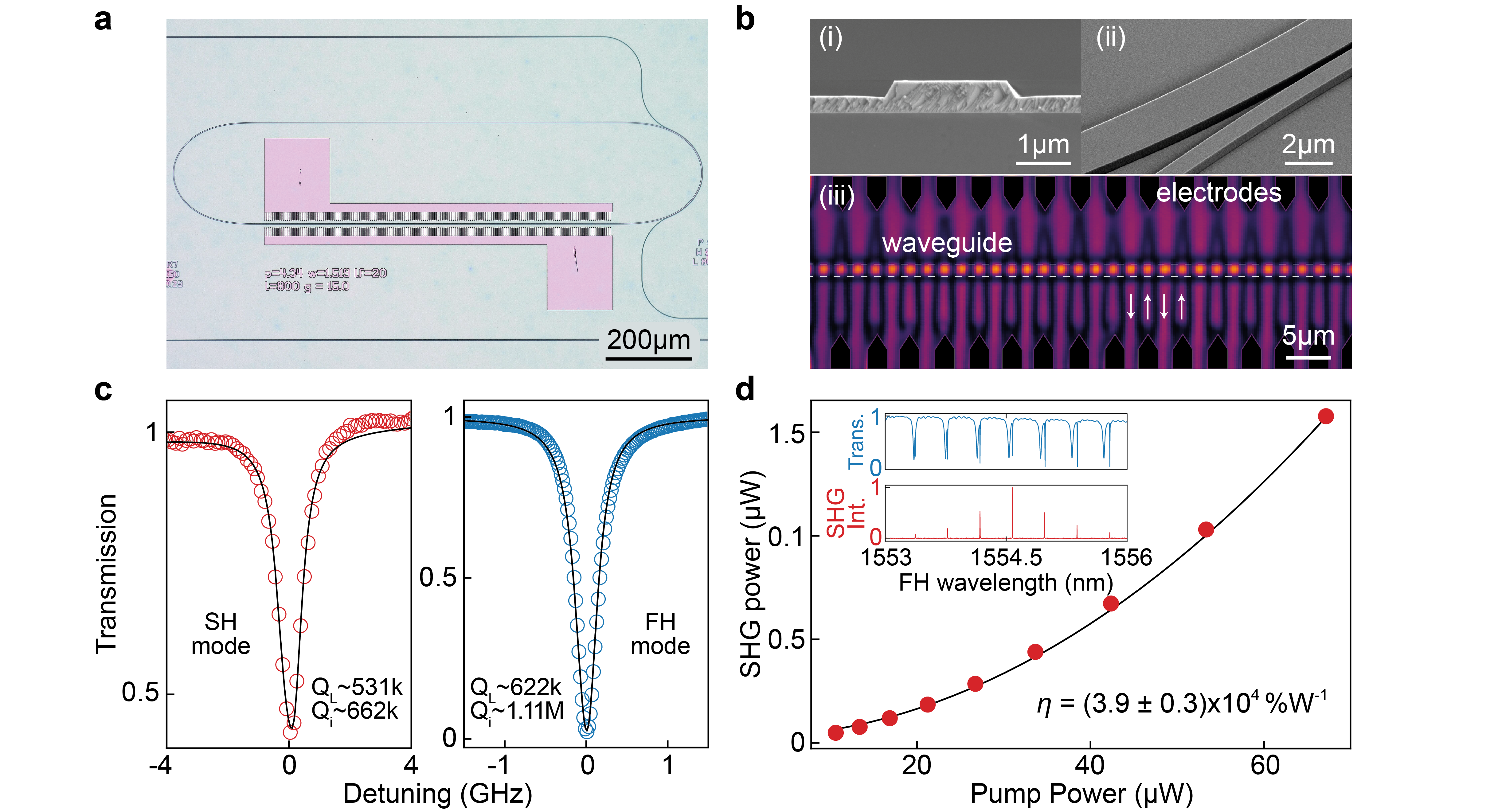}
    \caption{\textbf{Classical characterization of PPLT racetrack resonator.} \textbf{a,} Optical micrograph of PPLT racetrack resonator. \textbf{b,} SEM images of (i) waveguide cross-section and (ii) ring-bus waveguide coupling region, and (iii) SHG confocal microscopy image of the periodically poled region. \textbf{c,} Normalized transmission spectra of the SH and FH modes along with their respective quality factors. \textbf{d,} On-chip SHG power as a function of pump power with the normalized conversion efficiency of (3.9 $\pm$ 0.3) $\times$ 10$^4$ $\%$W$^{-1}$. Insets: PPLT racetrack transmission around FH wavelength showing 50 GHz free spectral range (FSR) and normalized SHG spectrum showing the optimal phase-matching wavelength.}
    \label{Fig3}
\end{figure*}

\section{Results}

\paragraph{PPLT waveguide source}
The nanophotonic waveguides are designed on a 310-nm-thick \textit{x}-cut TFLT platform, with a top width of 2 \textmu m and an etch depth of 200 nm. The poling period (3.24 \textmu m) enables phase matching between the fundamental transverse-electric modes at 1560 nm and 780 nm (see the design details in Methods). A 3-mm-long periodically poled section is made by electrical poling (Fig.\ref{Fig1}a), followed by electron-beam lithography and dry etching to pattern the waveguides in the poling region and cladding with silicon dioxide (Fig.\ref{Fig1}b, see fabrication details in Methods). 

We characterize the nonlinear efficiency of the PPLT waveguide by measuring its SHG response (bottom inset in Fig.\ref{Fig1}c).
Pumping at 1566 nm, the on-chip SHG power scales quadratically with the pump power, where a linear fitting slope of 2 in the log-log plot confirms operation in the non-depletion regime (Fig.\ref{Fig1}c). The normalized SHG conversion efficiency is calculated to be 841 $\pm$ 15 $\%$W$^{-1}$ cm$^{-2}$. We further map the phase-matching function by sum-frequency generation (SFG) (top inset in Fig.\ref{Fig1}c). The anti-diagonal pattern confirms that phase matching is maintained over a broad range of signal–idler wavelength combinations, consistent with a wide phase-matching bandwidth. Finally, the phase-matching wavelength shifts linearly with temperature at a rate of 0.28\,$\pm$\,0.06\,nm/\textdegree C (Fig.\,\ref{Fig1}d), allowing thermal tuning of the nonlinear source.

To produce SPDC photon pairs, we pump the PPLT waveguide with a continuous-wave (CW) laser at its second-harmonic (SH) phase-matching wavelength of 783 nm (Fig.\ref{Fig2}a). 
We measure the spectral correlation of the photon pairs by reconstructing the joint spectral intensity (JSI) (Fig.\ref{Fig2}b). 
The continuous anti-diagonal line shows the broadband photon-pair generation over 100 nm in the telecom band with strong frequency correlation. 
We further measure the pair generation rate (PGR) and CAR of the photon-pair source. 
The photon-counting rates for each channel and their coincidence histograms are recorded at different pump powers, from which the CAR and on-chip PGR are extracted (Fig.\ref{Fig2}c). 
The fitting line of PGR illustrates the linear relation to the pump power as expected for the SPDC process. 
The on-chip pair generation efficiency is calculated to be 2.1 $\pm$ 0.4 GHz mW$^{-1}$. 
The maximum CAR we observed is (3.8 $\pm$ 0.7) $\times$ 10$^5$ at a pump power of ~1 nW (PGR $\approx$ 8 kHz). 
As the pump power increases, the CAR decreases due to increased multi-pair events. 
At our highest on-chip pump power of 3 \textmu W, we perform heralded second-order correlation ($g_\mathrm{H}^{(2)}(\tau)$) measurement and obtain $g_\mathrm{H}^{(2)}(0)$ of 0.0018 $\pm$ 0.0002 (Fig.\ref{Fig2}d), showing that the operation is deep in the single-photon regime. We further perform Franson-type two-photon interferometry at the same power and achieve interference visibility of 98.9 $\pm$ 0.5 $\%$ (Fig.\ref{Fig2}e), verifying high fidelity time-energy entanglement.

\begin{figure*}[t]
    \centering
    \includegraphics[width = \textwidth]{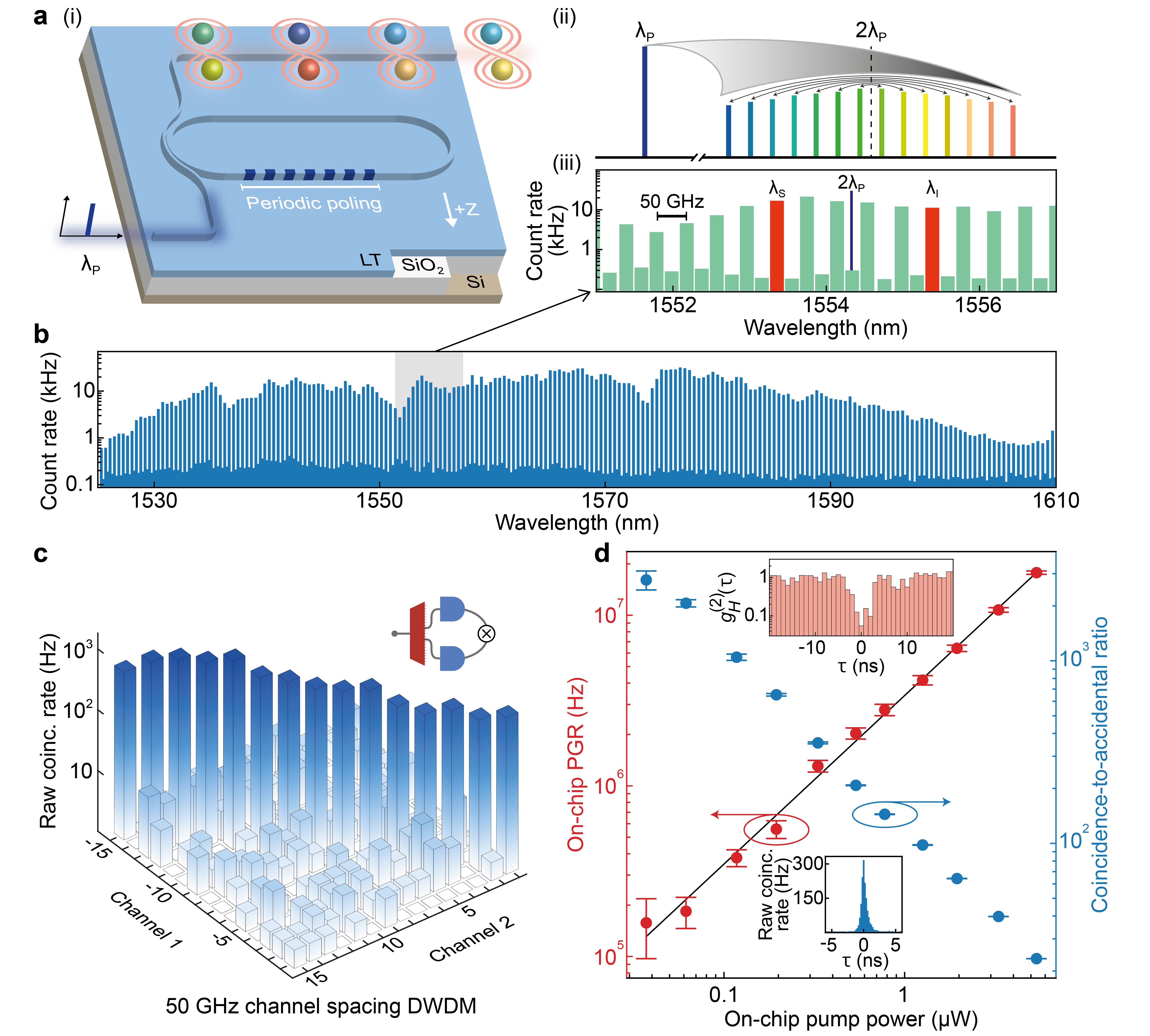}
    \caption{\textbf{Non-classical characterization of PPLT racetrack resonator.} \textbf{a,} (i) Schematic illustration of cavity enhanced SPDC process in TFLT racetrack resonator. (ii) The pump at wavelength of $\lambda_p$ gives rise to comb-like discrete photon pairs centered near 2$\lambda_p$. (iii) Zoomed-in section of a measured quantum frequency comb indicating the 50 GHz comb line spacing and the signal-idler pair chosen for the measurements. \textbf{b,} Measured quantum frequency comb in the PPLT racetrack. \textbf{c,} Joint spectral intensity of the photon pairs, reconstructed from correlations across 30 DWDM channels with 50 GHz frequency spacing, displaying strong frequency correlation. \textbf{d,} On-chip PGR (red) with pair generation efficiency of (3.3 $\pm$ 0.4) GHz mW$^{-1}$ and CAR (blue) as a function of on-chip pump power. Insets: Top - heralded second-order correlation measurement with the self correlation as a function of time delay, resulting in an anti-bunching dip $g_\mathrm{H}^{(2)}(0)$ of 0.06 $\pm$ 0.01. Bottom - raw coincidence histogram recorded at the highest pump power.}
    \label{Fig4}
\end{figure*}

\paragraph{PPLT microcavity source}

To improve spectral brightness, we integrate the PPLT waveguides in racetrack resonators to realize cavity-enhanced SPDC. The racetrack resonators used in this study are air-clad racetrack resonators fabricated on a 575-nm-thick TFLT with an etch depth of 300 nm and a waveguide width of 2 \textmu m (Fig.\ref{Fig3}a,b). 
One arm of the racetrack is periodically poled with a period of 4.32 \textmu m and a length of 800 \textmu m. 
We measure the transmission of the PPLT racetrack resonator around its fundamental-harmonic (FH) and SH wavelengths (Fig.\ref{Fig3}c) and extract the loaded quality factors of $6.22\times10^5$ and $5.31\times10^5$ and the intrinsic quality factors of $1.11\times10^6$ and $6.62\times10^5$, respectively. 
We measure the SHG spectrum (bottom inset in Fig.\ref{Fig3}d) and the on-chip SHG power as a function of on-chip pump power (Fig.\ref{Fig3}d). High efficiency SHG is achieved when both FH and SH are at resonant wavelengths (double resonance condition).
At 1554.58 nm, we measured the SHG power as a function of pump power, and extracted a normalized conversion efficiency of (3.9 $\pm$ 0.3) $\times$ 10$^4$ $\%$W$^{-1}$.


To generate SPDC photon pairs from the PPLT racetrack resonators, the device is pumped with a CW laser at SH resonant wavelength ($\sim$777 nm). 
The racetrack resonator enhances multiple phase-matched modes simultaneously, producing a broadband QFC of correlated photon pairs \cite{kuklewicz2006time} (Fig.\ref{Fig4}b). 
We measure the QFC spectrum by sweeping the tunable bandpass filter (TBPF) and collecting the singles counts using a superconducting nanowire single-photon detector (SNSPD). 
The QFC, having equally spaced comb lines with the spacing corresponding to the free spectral range (FSR) of the racetrack resonator (50.5 GHz), spans from 1525 nm to 1610 nm (limited by our detection wavelength range). 
We then reconstruct the JSI of the QFC using dense wavelength-division multiplexing (DWDM) with 50 GHz ITU grid (Fig.\ref{Fig4}c), which yields anti-diagonal points, indicating strong frequency correlation between the discrete and bright signal-idler pairs. 

We select one nondegenerate signal-idler pair to characterize the nonclassical correlations. The signal (CH30) and idler photons (CH27.5) are separated using a DWDM with a full-width at half-maximum (FWHM) of 0.3 nm. 
We measure the on-chip PGR and CAR as a function of on-chip pump power (Fig.\ref{Fig4}d). 
Similar to the PPLT waveguide, the PGR shows a linear relation to the pump power. 
We estimate the pair generation efficiency to be (3.3 $\pm$ 0.4) GHz mW$^{-1}$ for the selected frequency modes and achieve the highest CAR of (2.8 $\pm$ 0.2) × 10$^3$ at 37 nW power (PGR = 155 kHz). 
The spectral brightness of the photon pairs from the racetrack resonator is estimated to be 11 GHz mW$^{-1}$ GHz$^{-1}$ (or 1.3 THz mW$^{-1}$ nm$^{-1}$), which is calculated by normalizing the pair generation efficiency to the FWHM of the FH mode resonance ($\sim$2.5 pm). 
We also measure the heralded second-order correlation as a function of time delay, where a distinct anti-bunching dip is observed at $\tau$ = 0 (top inset in Fig.\ref{Fig4}d). The $g_\mathrm{H}^{(2)}(0)$ is measured to be 0.06 $\pm$ 0.01 at a pump power of 5.4 \textmu W, verifying that the measurements are operated in the non-classical regime.

\section{Discussion}
With the emergence of high-quality TFLT wafers and rapid advances in nanofabrication and electrical poling techniques, PPLT has recently attracted significant interest for nonlinear integrated photonics. 
As summarized in Supplementary Table S1, a diverse range of SHG devices has been demonstrated on the TFLT platform, spanning different geometries and performance regimes.

For microresonator-based devices, cavity properties play a central role in determining the nonlinear conversion efficiency and spectral characteristics. As compared in Table \ref{tab:table2}, our PPLT racetrack resonator demonstrates competitive performance, with spectral brightness exceeding 10 GHz mW$^{-1}$ GHz$^{-1}$. The measured quantum frequency comb spectrum reveals an asymmetric spectral envelope, with the photon count rate decreasing more rapidly at shorter wavelengths (Fig. \ref{Fig4}b). This behavior is attributed to the wavelength-dependent coupling condition, where the device transitions into the undercoupled regime more rapidly on the short-wavelength side (see Supplementary Fig. S1). These observations highlight the importance of precise dispersion and coupling engineering for achieving uniform and broadband comb generation.

From a fabrication perspective, we observe partial reversion of periodically poled domains after the etching and PECVD processes. This effect is likely related to the relatively low Curie temperature of LT \cite{tormo2019low}, where localized heating during etching may approach or exceed the ferroelectric transition temperature (600–700$^\circ$C), leading to domain instability. This sensitivity suggests that PPLT devices require careful thermal management during subsequent processing steps, including plasma etching, cladding deposition, and annealing. Further optimization of fabrication processes will be essential to ensure long-term domain stability and reproducible device performance.

Beyond discrete-variable photon-pair generation, PPLT microresonators also offer a promising route toward cavity-enhanced squeezed light generation for continuous-variable quantum technologies. Compared to photon-pair generation, squeezed-light generation typically requires higher intracavity power and long-term operational stability \cite{arge2025demonstration}. In this context, the weak photorefractive response and high optical damage threshold of LT provide key advantages, enabling stable operation in regimes that are challenging for other $\chi^{(2)}$ platforms.

Looking forward, the combination of wafer-scale TFLT fabrication, strong $\chi^{(2)}$ nonlinearity, high refractive-index contrast, and electro-optic functionality positions LT as a highly versatile platform for integrated quantum photonics. The ability to co-integrate nonlinear sources, high-speed modulators, wavelength multiplexers, and reconfigurable photonic circuits on a single chip opens a pathway toward large-scale quantum photonic systems. 

\renewcommand{\arraystretch}{1.3}
\begin{table*}[t]
\centering
\caption{\label{tab:table2} \textbf{Comparison of microresonator based photon-pair sources} in terms of integrated material platforms, nonlinear processes, pair-generation rate (PGR), loaded Q-factor, brightness, and heralded second-order correlation.}
\renewcommand{\arraystretch}{1.2} 
\begin{ruledtabular}
\begin{tabular}{ccccccc}
 Ref. & Material & Process & \makecell[c]{PGR@1mW \\ (Hz)} & Q-factor & \makecell[c]{Brightness@1mW \\ (Hz/GHz)}& $g_\mathrm{H}^{(2)}(0)$ \\
\hline
\cite{ma2020ultrabright} & TFLN & SPDC & 2.7 $\times$ 10$^9$ & 1.0 $\times$ 10$^5$ & 1.4 $\times$ 10$^9$ & 0.008 \\
\cite{chen2025efficient}& TFLN & SPDC & 4.0 $\times$ 10$^7$ & 1.3 $\times$ 10$^5$ & 2.6 $\times$ 10$^7$& \\
\cite{zhao2022ingap}& InGaP & SPDC & 2.7 $\times$ 10$^{10}$ & 1.1 $\times$ 10$^5$ & 1.5 $\times$ 10$^{10}$ & \\
\cite{guo2017parametric}& AlN & SPDC & 5.8 $\times$ 10$^6$ & 2.0 $\times$ 10$^5$ & 6.0 $\times$ 10$^6$ & 0.088  \\
\cite{steiner2021ultrabright}& AlGaAs & SFWM & 2.0 $\times$ 10$^{10}$ & 1.2 $\times$ 10$^6$ & 2.0 $\times$ 10$^{11}$ & 0.004 \\
\cite{ma2017silicon}& Si & SFWM & 1.5 $\times$ 10$^8$ & 9.2 $\times$ 10$^4$ & 1.6 $\times$ 10$^8$ & 0.0053 \\
\cite{yasui2025narrowband}& Si & SFWM & 1.3 $\times$ 10$^9$ & 6.5 $\times$ 10$^5$ & 3.9 $\times$ 10$^9$ & \\
\cite{chen2024ultralow}& SiN & SFWM & 3.0 $\times$ 10$^7$ & 5.0 $\times$ 10$^6$ & 1.2 $\times$ 10$^9$ & 0.0037 \\
\cite{li2025down}& SiN & SPDC & 5.3 $\times$ 10$^5$ & 1.4 $\times$ 10$^7$ & 6.0 $\times$ 10$^7$ & \\
\cite{luo2026scalable}& SiN & SFWM & 9.3 $\times$ 10$^7$ & 1.4 $\times$ 10$^6$ & 5.8 $\times$ 10$^9$ & \\
\cite{rahmouni2024entangled}& 4H-SiC & SFWM & 3.1 $\times$ 10$^5$ & 8.0 $\times$ 10$^5$ & 1.3 $\times$ 10$^6$ & \\
\cite{li2024integrated}& 3C-SiC & SPDC & 9.6 $\times$ 10$^5$ & 2.5 $\times$ 10$^4$ & 1.25 $\times$ 10$^5$ & 0.0007\\
\hline
This work & TFLT & SPDC & 3.3 $\times$ 10$^9$ & 6.2 $\times$ 10$^5$ & 1.1 $\times$ 10$^{10}$ & 0.06\\
\end{tabular}
\end{ruledtabular}
\renewcommand{\arraystretch}{1} 
\end{table*}

\section*{Methods}

\paragraph{Device design}

The PPLT nanophotonic waveguides are designed on a 310-nm-thick $x$-cut TFLT platform with a top width of 2 \textmu m, an etch depth of 200 nm, and 1.5 \textmu m silicon dioxide cladding. The PPLT racetrack resonators are designed on a 575-nm-thick $x$-cut TFLT platform, with a 2 \textmu m waveguide width, 300 nm etch depth, and air cladding. The effective indices ($n$) and mode distribution of FH and SH modes are numerically simulated using a finite difference mode solver (Ansys Lumerical MODE). The poling periods for both the waveguide ($\it\Lambda$ = 3.24 \textmu m) and racetrack resonator ($\it\Lambda$ = 4.32 \textmu m) are designed to attain the quasi-phase matching between the 1560 nm TE$_{00}$ and 780 nm TE$_{00}$ modes. The poling period is calculated by $\it\Lambda = \lambda_{\text{FH}}/\,2(n_{\text{SH}}-n_{\text{FH}})$, where $\lambda_{\text{FH}}$ is the FH wavelength and $n_{\text{SH/FH}}$ are the effective indices of SH and FH modes.

\paragraph{Device fabrication}

The PPLT straight waveguide and racetrack resonator are fabricated on 310 nm and 575 nm $x$-cut TFLT (Omedasemi), respectively. For the poling process, a buffer layer of 15 nm hafnium dioxide (HfO$_2$) is deposited using atomic layer deposition (ALD). The poling electrodes are patterned by electron beam lithography (EBL), followed by the e-beam evaporation (Nickel - 100 nm) and lift-off. The LT film is periodically poled using a series of high voltage pulses, and the quality of poling is assessed by SHG confocal microscopy (Fig.\ref{Fig1}b(ii), Fig.\ref{Fig3}b(iii)). After poling, the buffer layer is removed. The waveguides are patterned using EBL, followed by Ar$^+$ etching in inductively coupled plasma reactive ion-etching (ICP RIE) with maN 2405 as the etching mask. The redeposition generated by dry etching is removed using an alkaline solution. The residual resist is then removed by oxygen plasma. Finally, a 1.5 \textmu m thick silicon dioxide film is deposited on top of the chip using plasma-enhanced chemical vapor deposition (PECVD). The PPLT racetrack resonator chip remains air-cladded. 

\paragraph{Classical characterization}

When measuring SHG in PPLT devices, a tunable telecom laser, serving as the pump, is coupled to the chip using a lensed fiber after passing through a tunable attenuator to adjust the pump power, a 90:10 fiber-based beam splitter (BS), and a polarization controller. The 10$\%$ arm output of the BS is for pump power monitoring. The output light from the chip is coupled through another lensed fiber and connected to a power meter after passing through a 775/1550 nm wavelength division multiplexer (WDM). The SHG spectrum (Inset - Fig.\ref{Fig1}c, Fig.\ref{Fig3}d) is characterized by sweeping the telecom laser and recording the power-meter output signal using a data-acquisition system. Then, the pump wavelength is fixed at the SHG peak wavelength, and the corresponding SHG is recorded with varied pump power. The SHG versus pump power graph is then plotted after subtracting the coupling losses (Fig\ref{Fig1}c, Fig.\ref{Fig3}d). 

To obtain the phase-matching function of the PPLT nanophotonic waveguide using SFG (Fig.\ref{Fig1}c [top inset]), two tunable CW telecom lasers are used to pump the waveguide after combining through a 50/50 beam splitter, and the SFG power is measured after filtering the pumps. The lasers are swept synchronously to reconstruct the phase-matching function.

\paragraph{Joint spectral intensity}

To generate photon pairs, the chips are pumped with a CW visible laser at the phase-matching wavelength. The pump is coupled to the chip after passing through a short-pass filter (SPF), a 90:10 fiber-based BS and a polarization controller. After the chip, the pump residue is filtered out by 775/1550 nm WDM and long-pass filters (LPF), and the photon pairs are detected using SNSPD after passing through a 50:50 fiber-based BS and TBPF (for nanophotonic waveguide device) or a DWDM (for racetrack resonator device). The detected signals are processed using a time-to-digital converter (TDC) to record the coincidence rate and singles count rate. The coincidence rates are recorded by sweeping the TBPFs/DWDM-channels synchronously, which is then used to reconstruct the JSI.

\paragraph{Pair generation rate and CAR measurements}

The photon pairs from the chip are split into two paths using a 50:50 fiber-based BS (for the nanophotonic waveguide) or a DWDM with channel spacing of 50 GHz and channel bandwidth of 0.3 nm (for the racetrack resonator) and are detected using SNSPDs. For the racetrack resonator device, DWDM CH30 (1553.33 nm) and CH27.5 (1555.34 nm) are chosen as the signal and idler channels, respectively.\\
The PGR in Fig.\ref{Fig2}c is calculated by \cite{zhao2020high}
\begin{gather*}
    \text{PGR}_{\mathrm{BS}} = \frac{N_{\mathrm{S}} N_{\mathrm{I}}}{2N_{\mathrm{SI}}}
\end{gather*}
The PGR in Fig.\ref{Fig4}d is calculated by \cite{chen2024ultralow}
\begin{gather*}
    \text{PGR}_{\mathrm{DWDM}} = \frac{N_{\mathrm{S}} N_{\mathrm{I}}}{N_{\mathrm{SI}}}
\end{gather*}
where $N_{\mathrm{SI}}$ is the coincidence rate within the coincidence window, and $N_{\mathrm{S}}$ and $N_{\mathrm{I}}$ are the singles count rates for signal and idler. \\
The CAR is calculated by $N_{\mathrm{CC}}$/$N_{\mathrm{AC}}$, where $N_{\mathrm{CC}}$ is the total coincidence count within the coincidence window, and $N_{\mathrm{AC}}$ is the total accidental counts estimated by averaging the background counts.

\paragraph{Heralded second-order correlation}

Heralded $g_\mathrm{H}^{(2)}(\tau)$ measurements are performed via a three-fold coincidence detection scheme. The signal channel serves as the heralding channel (H), whereas the idler photons are split using a 50/50 BS (channels A and B). A virtual delay ($\tau$) is introduced to channel B through TDC. The triple coincidence ($N_{\mathrm{HAB}}$) and the double coincidence between the heralding channel and the other two channels ($N_{\mathrm{HA}}$ and $N_{\mathrm{HB}}$) are recorded.\\
The value of $g_\mathrm{H}^{(2)}(\tau)$ in Fig.\ref{Fig2}d is calculated by \cite{zhao2020high} 
\begin{gather*}
    g_\mathrm{H}^{(2)}(\tau)_{\rm BS} = \frac{N_\mathrm{H}.N_{\mathrm{HAB}}(\tau)}{2\,N_{\mathrm{HA}}.N_{\mathrm{HB}}(\tau)}
\end{gather*}
The value of $g_H^{(2)}(\tau)$ in Fig.\ref{Fig4}d is calculated by \cite{steiner2021ultrabright} 
\begin{gather*}
    g_\mathrm{H}^{(2)}(\tau)_{\rm DWDM} = \frac{N_\mathrm{H}.N_{\mathrm{HAB}}(\tau)}{N_{\mathrm{HA}}.N_{\mathrm{HB}}(\tau)}
\end{gather*}

\paragraph{Franson interferometry}

The SPDC photon pairs are expected to have time-energy entanglement, and this can be verified by a Franson-type two-photon interference experiment. The output photons after passing through the LPFs are connected to a Mach-Zehnder interferometer (MZI). A fiber stretcher is employed in one of the MZI arms to introduce the phase delay. One of the outputs is connected to a DWDM, and a single-idler channel pair is coupled to the SNSPDs. A sinusoidal curve of the measured coincidence rate is shown due to the phase modulation actuated by a fiber stretcher. The raw interference visibility is calculated as
\begin{gather*}
    V = \frac{N_{\mathrm{CC-max}}\,-\,N_{\mathrm{CC-min}}}{N_{\mathrm{CC-max}}\,+\,N_{\mathrm{CC-min}}}
\end{gather*}
where $N_{\mathrm{CC-max}}$ and $N_{\mathrm{CC-min}}$ are the measured values of maxima and minima of the coincidence rate.

\section*{Acknowledgements}

This research was supported by A*STAR under its NSTIC program (M25W2NS002), NRF under its NQFF program (W24Q3D0003), Singapore Ministry of Education under the Academic Research Fund Tier 3 (MOET32024-0009) and Tier  1 (FY2025), and Centre for Quantum Technologies Funding Initiative (S24Q2d0009).

\section*{Competing interests}
The authors declare no competing interests.

\bibliography{ref}

\end{document}